\pdfoutput=1
\documentclass[prl,twocolumn,superscriptaddress]{revtex4-1} 

\usepackage{graphicx}
\usepackage{amsmath}
\usepackage{amssymb}
\usepackage[colorlinks=true, linkcolor=blue, urlcolor=blue,
            citecolor=blue]{hyperref}
\usepackage{bm}
\usepackage[utf8x]{inputenc}

\begin{document}
\title{Quantum spin torque driven transmutation of antiferromagnetic Mott insulator}

\author{Marko D.~Petrovi\'c}
\affiliation{Department of Physics and Astronomy, University of Delaware, Newark DE 19716, USA}

\author{Priyanka Mondal}
\affiliation{Department of Physics and Astronomy, University of Delaware, Newark DE 19716, USA}

\author{Adrian E.~Feiguin}
\affiliation{Department of Physics, Northeastern University, Boston, MA 02115, USA}

\author{Branislav K.~Nikoli\'c}
\email{bnikolic@udel.edu}
\affiliation{Department of Physics and Astronomy, University of Delaware, Newark DE 19716, USA}

\begin{abstract}
	The standard model of spin-transfer torque (STT) in  antiferromagnetic spintronics considers exchange of angular momentum between quantum spins of flowing electrons and noncollinear-to-them localized spins  treated as {\em classical} vectors. These vectors are assumed to realize N\'{e}el order in equilibrium,  $\uparrow \downarrow \ldots \uparrow \downarrow$, and their STT-driven  dynamics is described by the Landau-Lifshitz-Gilbert (LLG) equation. However, many experimentally employed materials (such as archetypal NiO) are strongly electron-correlated antiferromagnetic Mott insulators (AFMI) where localized spins form a ground state quite different from the unentangled N\'{e}el state $|\!\! \uparrow \downarrow \ldots \uparrow \downarrow \rangle$.  The true ground state is entangled by quantum spin fluctuations, leading to expectation value of all localized spins being {\em zero}, so that LLG dynamics of classical vectors of fixed length rotating due to STT cannot even be initiated. Instead, a fully quantum treatment of {\em both} conduction electrons and localized spins is necessary to capture exchange of spin angular momentum between them, denoted as {\em quantum STT}. We use a recently developed time-dependent density matrix renormalization group approach to quantum STT to predict how injection of a spin-polarized current pulse into a normal metal layer coupled to AFMI overlayer via exchange interaction and possibly small interlayer hopping---which mimics, e.g., topological-insulator/NiO bilayer employed  experimentally---will induce nonzero expectation value of AFMI localized spins. This new nonequilibrium phase is a spatially inhomogeneous ferromagnet with zigzag profile of localized spins. The total spin absorbed by AFMI increases with electron-electron repulsion in AFMI, as well as when the two layers do not exchange any charge.  
\end{abstract}

\maketitle

{\em Introduction}.---The emergence of antiferromagnetic spintronics~\cite{Baltz2018,Jungwirth2016,Zelezny2018,Jungfleisch2018} has elevated 
antiferromagnetic (AF) insulators (AFIs) and metals into active elements of spintronic devices. They exhibit dynamics of their localized spins at a much higher frequencies, reaching THz~\cite{Jungfleisch2018}, when compared to ferromagnetic spintronics. Furthermore, the absence of net magnetization forbids any  stray magnetic fields, making them largely insensitive to perturbations by external fields. They also exhibit magnetoresistance effects~\cite{Manchon2017,Baldrati2018} enabling electric readout of changes in the orientations of their localized spins.

Basic spintronic phenomena like spin-transfer torque (STT)~\cite{Chen2018,Moriyama2018,Gray2019,Wang2019}---where spin angular momentum is exchanged between flowing conduction electrons and noncollinear-to-them~\cite{Ralph2008} localized spins---and spin pumping~\cite{Vaidya2020}---where precessing localized spins pump pure spin current in the absence of any bias voltage---have been demonstrated recently using different AF materials.  The theoretical description~\cite{Nunez2006,Haney2008,Xu2008,Stamenova2017,Gomonay2010,Hals2011,Cheng2014,Cheng2015,Saidaoui2014,Dolui2020} of these phenomena invariably assumes that localized magnetic moments on two sublattices of the AF material, $\mathbf{M}_i^A$ and $\mathbf{M}_i^B$,  are {\em classical} vectors with net zero total magnetization in equilibrium due to assumed N\'{e}el classical ground state (GS), $\uparrow \downarrow \ldots \uparrow \downarrow$. Out of equilibrium, the dynamics of such classical vectors of {\em fixed} length is described by the Landau-Lifshitz-Gilbert (LLG)  equation~\cite{Evans2014}. The STT is typically  introduced into the LLG equation either as a phenomenological term~\cite{Gomonay2010,Hals2011,Cheng2014,Cheng2015}, or it is calculated microscopically by using  steady-state single-particle quantum transport formalism applied to model~\cite{Nunez2006,Haney2008,Saidaoui2014} or  first-principles~\cite{Xu2008,Stamenova2017,Dolui2020} Hamiltonians of AF materials. Recently STT~\cite{Suresh2020a} from time-dependent single-particle quantum transport formalism~\cite{Gaury2014} has been coupled~\cite{Petrovic2018} to the LLG equation, capturing additional quantum effects like electronic spin pumping by moving $\mathbf{M}_i^A(t)$ and $\mathbf{M}_i^B(t)$ and the corresponding enhanced damping on them, but this remains conventional~\cite{Ralph2008}  {\em quantum}-for-electrons--{\em classical}-for-localized-spins theory of STT.

\begin{figure}
	\includegraphics[width=0.49\textwidth]{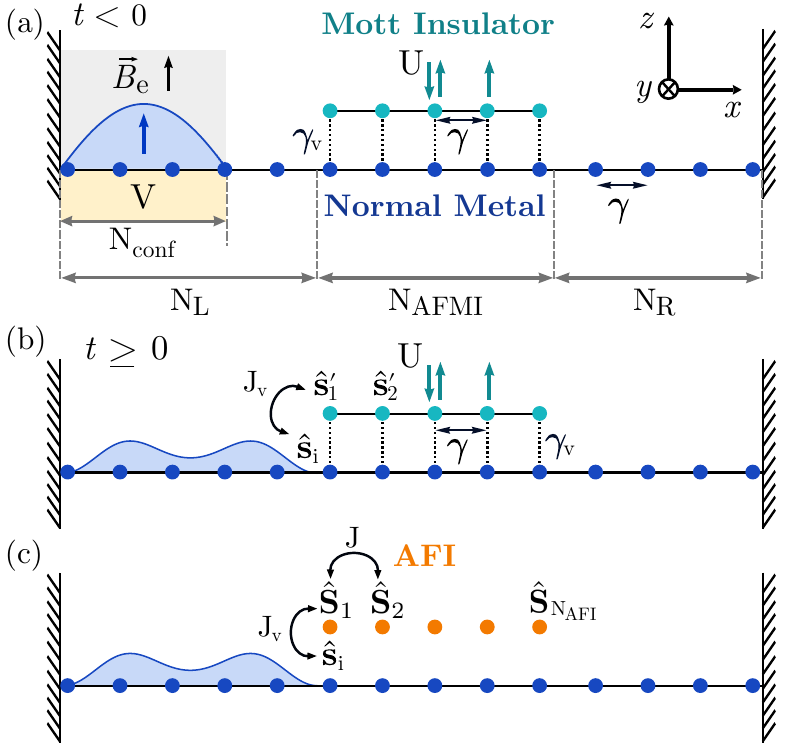}
	\caption{\label{fig:fig1}
	(a) Schematic view of a ``bilayer''~\cite{Wang2019} for tDMRG calculations. The 1D TB chain (blue dots) of $N=N_\mathrm{L}+N_\mathrm{AFMI}+N_\mathrm{R}=92$ sites, with intrachain hopping $\gamma$, models the NM surface (such as that of Bi$_2$Se$_3$ in experiment of Ref.~\cite{Wang2019}) through which spin-polarized current pulse is injected. The pulse exerts quantum STT on a Hubbard chain of $N_{\rm AFMI} = 12$ sites with the on-site Coulomb repulsion $U$, modeling the surface of strongly electron-correlated AFMI (such as that of NiO in Refs.~\cite{Chen2018,Moriyama2018,Gray2019,Wang2019}).  The electronic spins in two chains interact via interchain exchange interaction $J_{\rm v}$, and we consider both $\gamma_{\rm v}=0$ and $\gamma_{\rm v} \neq 0$ interchain hopping where the latter mimics possible hybridization of NM and AFMI via evanescent wavefunctions~\cite{Dolui2020}. For times $t < 0$, $N_{\rm e} = 12$ noninteracting electrons are confined by potential $V$ within $N_{\rm conf} = 25$ sites of the L lead  	(composed of $N_\mathrm{L}=40$ sites), as well as spin-polarized by an external magnetic field $\mathbf{B}_{\rm e}$ pointing along the $z$-axis. Concurrently, $N_{\rm e}^{\rm AFMI} = 12$ electrons half-fill the AFMI chain. For times $t \ge 0$, $V$ and $\mathbf{B}_{\rm e}$ are removed, so that electrons propagate as spin-polarized  current pulse from the L to the R lead, as animated in the movie in the SM~\cite{sm}. In panel (c) we replace AFMI from (a) and (b) with AF quantum Heisenberg chain where spin-$\frac{1}{2}$ operators reside on each (orange) site and interact via $J=4\gamma^2/U$ in Eq.~\eqref{eq:afichain} with no electrons allowed within this chain.}
\end{figure}

However, AFIs employed in spintronics experiments are typically strongly electron-correlated transition metal oxides due to narrow $d$ bands. For example, widely used~\cite{Chen2018,Moriyama2018,Gray2019,Wang2019,Baldrati2018} NiO shares features of both Mott and charge-transfer insulators~\cite{Karolak2010,Lechermann2019,Zhang2018,Gillmeister2020}. Due to quantum (or zero-point) spin fluctuations~\cite{Singh1990,Humeniuk2019,Kamra2019}, the AF GS  is highly {\em entangled}~\cite{Humeniuk2019,Roscilde2005,Vafek2017,Christensen2007}, which results in {\em zero} expectation value of all localized spins, $\mathbf{S}_i = 0$ ($\mathbf{M}_i \propto \mathbf{S}_i = 0$). Thus, conventional~\cite{Ralph2008} STT $\propto \mathbf{s}_i \times \mathbf{S}_i=0$ due to injected nonequilibrium electronic spin density $\mathbf{s}_i$  {\em cannot} be initiated because $\mathbf{S}_i(t=0) \equiv 0$. Even if \mbox{$|\mathbf{S}_i(t=0)| \neq 0$} is provoked by spin-rotation-symmetry-breaking anisotropies~\cite{Stoudenmire2012} or impurities [see Supplemental Material (SM)~\cite{sm} for illustration], the LLG equation is inapplicable~\cite{Wieser2015,Wieser2016} because the length $|\mathbf{S}_i(t)| < |\mathbf{S}_i^\mathrm{N\acute{e}el}|$ will be changing in time,  with smaller value signifying higher entanglement (unobserved  quantum systems exhibit unitary evolution toward states of higher entanglement~\cite{Skinner2019}). Thus, both situations {\em necessitate} to describe localized spins {\em fully} quantum mechanically where their expectation values $\mathbf{S}_i(t)$ are calculated {\em only at the end}.

The entanglement in the AF GS leading to $\mathbf{S}_i = 0$ can be illustrated using an example~\cite{Fradkin2013,Essler2005} of a one-dimensional (1D)  AF quantum spin-$\frac{1}{2}$ Heisenberg chain 
\begin{equation}\label{eq:afichain}
\hat{H}_{\textrm{AFI}} = J \sum_{i = 1}^{N_{\textrm{AFI}}-1}
\hat{\mathbf{S}}_{i}\cdot \hat{\mathbf{S}}_{i+1},
\end{equation}
on $N_{\textrm{AFI}}$ sites. Here $\hat{S}_{i}^\alpha =\hat{I}_1 \otimes \ldots \otimes \frac{1}{2}\hat{\sigma}^\alpha \otimes \ldots \otimes \hat{I}_\mathrm{N_\mathrm{AFI}}$ acts nontrivially, as the Pauli matrix $\hat{\sigma}^\alpha$, only on the Hilbert space of site $i$; $\hat{I}_i$ is the unit operator; and  $J > 0$ is AF exchange interaction. The true GS is easy to write explicitly for small $N_\mathrm{AFI}$, such as for $N_\mathrm{AFI}=4$ we find
$|\mathrm{GS}\rangle  =  \frac{1}{\sqrt{12}} \big( 2|\!\!\uparrow \downarrow \uparrow \downarrow\rangle + 2|\!\!\downarrow \uparrow \downarrow \uparrow\rangle - |\!\!\uparrow\uparrow \downarrow  \downarrow \rangle - |\!\!\uparrow \downarrow \downarrow \uparrow  \rangle  - |\!\!\downarrow \downarrow \uparrow \uparrow \rangle -|\!\!\downarrow \uparrow \uparrow \downarrow \rangle \big)$. Its energy, $\langle \mathrm{GS}| \hat{H}_{\textrm{AFI}} | \mathrm{GS} \rangle = -2J$, is lower  than the energy of the unentangeled (i.e., direct-product) N\'{e}el state, $\langle \uparrow\downarrow
\uparrow\downarrow \!\!| \hat{H}_{\textrm{AFI}} |\!\! \uparrow\downarrow \uparrow\downarrow \rangle=-J$.  This is in sharp contrast to ferromagnets where quantum spin fluctuations are absent, and both classical $\uparrow \uparrow \ldots \uparrow \uparrow$ and its unentangled quantum counterpart 
$|\!\! \uparrow \uparrow \ldots \uparrow \uparrow \rangle$ are GS of the respective classical and quantum Hamiltonian [such as Eq.~\eqref{eq:afichain} with $J<0$]. This justifies~\cite{Wieser2015,Wieser2016} the picture of interacting classical $\mathbf{M}_i$ in spintronics~\cite{Ralph2008} and micromagnetics~\cite{Evans2014}, even as the size of the localized spin is reduced to that of a single electron spin. Conversely, in the case of 
many-body entangled~\cite{Humeniuk2019,Roscilde2005,Vafek2017,Christensen2007} AF GS, the quantum state of each localized spin subsystem {\em must} be described by the reduced density matrix, $\hat{\rho}_{i} = \mathrm{Tr}_\mathrm{other} \, |\mathrm{GS} \rangle \langle \mathrm{GS}|$, where partial trace is performed in the Hilbert subspace of all other localized spins $j \neq i$. The expectation value
\begin{equation}\label{eq:si}
\mathbf{S}_i \equiv \langle \hat{\mathbf{S}}_i \rangle =\mathrm{Tr}\,[ \hat{\rho}_i \hat{\mathbf{S}}_i], 
\end{equation}
is then identically zero vector, $\mathbf{S}_i=0$, on all sites (see the SM~\cite{sm}). The GS in the limit $N_\mathrm{AFI} \rightarrow \infty$ is computable by Bethe ansatz~\cite{Essler2005}, and its entanglement ensures \mbox{$\mathbf{S}_i=0$}. The entanglement in the GS of crystalline realization of a 
two-dimensional (2D) quantum Heisenberg antiferromagnet or antiferromagnetic Mott insulator (AFMI) realized with cold atoms on a square lattice has been detected by neutron scattering~\cite{Christensen2007} or optically~\cite{Mazurenko2017}, respectively, at ultralow temperatures.

In this Letter, we employ the emerging concept of quantum STT~\cite{Zholud2017,Mondal2019,Petrovic2020,Mitrofanov2020} where {\em both} conduction electrons and localized spins are treated fully quantum-mechanically to describe the exchange of spin angular momentum between them. This allows us to predict {\em nonequilibrium phase transition} of AFMI driven by {\em absorption} of spin angular momentum from spin-polarized current pulse injected into an adjacent normal metal (NM). To model such genuine quantum many-body problem, we evolve in time a nonequilibrium quantum state of NM/AFMI  system via very recently adapted~\cite{Petrovic2020} to quantum STT time-dependent density matrix renormalization group (tDMRG) approach~\cite{White2004,Feiguin2011,Daley2004,Paeckel2019}.

Our system geometry in Fig.~\ref{fig:fig1} consists of a NM modeled as 1D tight-binding (TB) chain, which is split into the left (L) and the right (R) leads sandwiching a central region. The conduction electron spins in the central region are exchange coupled to an AFMI chain modeled by Hubbard model with the on-site Coulomb repulsion $U$. The current pulse, carrying electrons initially spin-polarized in the direction perpendicular to the interface (i.e., along the $z$-axis in Fig.~\ref{fig:fig1}), is injected from the L lead into the central region of NM in order to initiate the AFMI dynamics via quantum STT.
\begin{figure}
	\includegraphics[width=\linewidth]{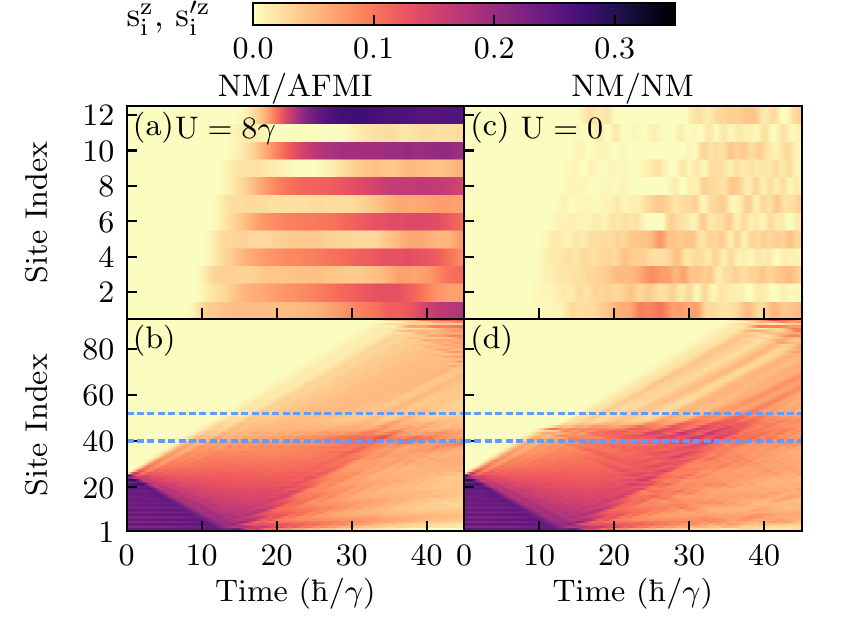}
	\caption{Spatio-temporal profiles of the $z$-component of  spin density within: (a) AFMI chain with the Coulomb repulsion $U = 8\gamma$; and (c) the same chain with $U=0$ acting as the second NM chain half-filled with electrons. In both panels $s_i^{\prime z}(t=0) \equiv 0$, so that only  $s_i^{\prime z}(t) \neq 0$ component is induced by current pulse spin-polarized along the $z$-axis  and flowing along the bottom NM chain in Fig.~\ref{fig:fig1}(b) whose $s_i^z$ profiles in panels (b) and (d) are driving the profiles in panels (a) and (c), respectively, via quantum STT. The dotted horizontal lines in (b) and (d) mark the boundaries between the leads and the central region of the NM chain in Fig.~\ref{fig:fig1}. The interchain exchange is $J_{\rm v} = 0.5\gamma$ and hopping $\gamma_{\rm v} = 0$ in Eq.~\eqref{eq:int}. All four panels, together with the corresponding electron densities, are animated in the movie in the SM~\cite{sm}.}\label{fig:fig2}
\end{figure}
Our geometry mimics recent experiment~\cite{Wang2019} on injection of current pulses into metallic surface of topological insulator Bi$_2$Se$_3$, which then exert spin torque on the surface of NiO overlayer covering Bi$_2$Se$_3$, except that in the experiment spin-orbit coupling polarizes injected electrons in the plane of the interface (i.e., along the $y$-axis in Fig.~\ref{fig:fig1}). Nevertheless, since singlet with $\mathbf{s}^\prime_i(t=0) \equiv 0$ on all sites of AFMI is rotationally invariant, the final spin state of AFMI driven by quantum STT will be the same for arbitrary spin-polarization of injected electrons.

\begin{figure}
	\includegraphics[width=\linewidth]{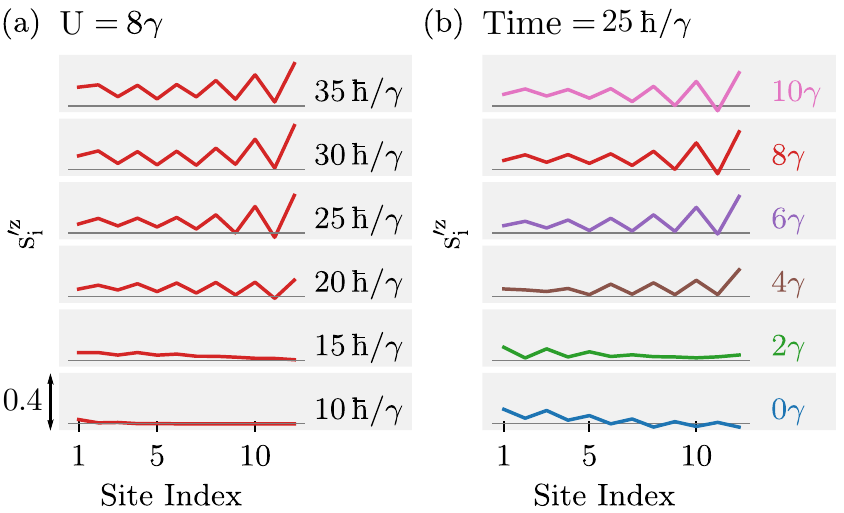}
	\caption{Spatial profile of the $z$-component $s_i^{\prime z}$ of spin density within AFMI chain in Fig.~\ref{fig:fig1}(b) driven by quantum STT from NM chain: (a) at different times using $U = 8\gamma$ in Eq.~\eqref{eq:hmott}; and (b) for different $U$ values at time $t=25 \hbar/\gamma$. The interchain exchange  is $J_{\rm v} = 0.5\gamma$ and hopping $\gamma_{\rm v} = 0$ in Eq.~\eqref{eq:int}.}\label{fig:fig3}
\end{figure}

\begin{figure}
	\includegraphics[width=\linewidth]{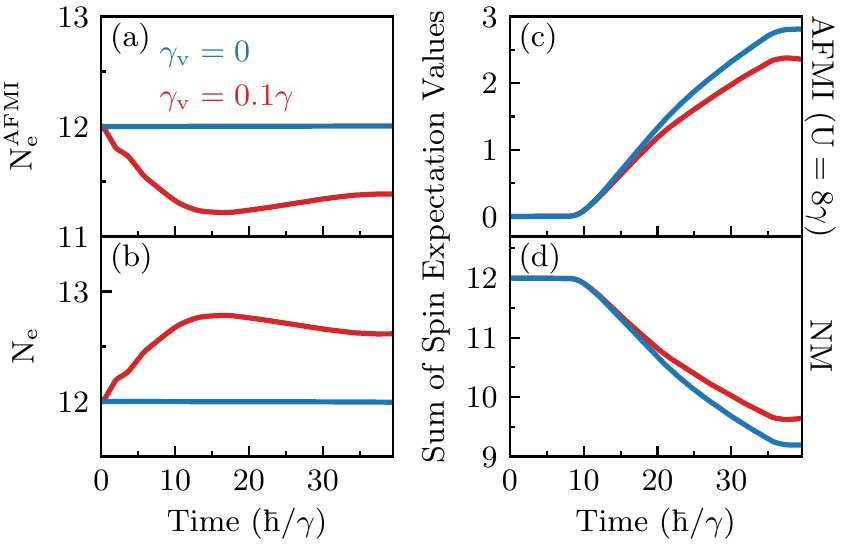}
	\caption{Time dependence of the total number of electrons within (a) AFMI and (b) NM chains in the setup of Fig.~\ref{fig:fig1}(b) for
		two different interchain hoppings \mbox{$\gamma_{\rm v} = 0$} (blue lines) and \mbox{$\gamma_{\rm v} = 0.1\gamma$} (red lines). Panels (c) and (d) 
		show the corresponding time dependence of the sum of the $z$-component of spin densities, $\sum_i s_i^{\prime z}$ and $\sum_i s_i^{z}$, respectively. The on-site Coulomb repulsion is $U=8\gamma$ [Eq.~\eqref{eq:hmott}] within the AFMI and interchain exchange interaction is $J_{\rm v} = 0.5\gamma$.}\label{fig:fig4}
\end{figure}

\begin{figure}
	\includegraphics[width=\linewidth]{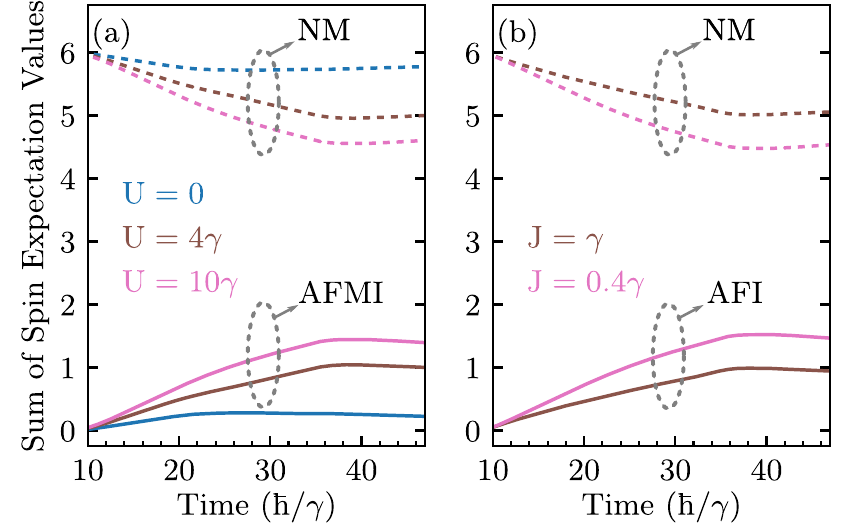}
	\caption{(a) Time evolution of the sum  of spin densities within NM chain $\sum_i s_i^z$ (dashed lines) and AFMI chain $\sum_i s_i^{\prime z}$ (solid lines) in the setup of Fig.~\ref{fig:fig1}(b) for different value of the  on-site Coulomb repulsion $U$ within the AFMI chain. For comparison, panel (b) plots the same information for the setup in Fig.~\ref{fig:fig1}(c) where AFMI is replaced by AF quantum spin-$\frac{1}{2}$ Heisenberg chain  with no electrons, so that solid lines are  $\sum_i S_i^z$ defined in Eq.~\eqref{eq:si}.  For each $U$ in (a), we set the corresponding intrachain exchange interaction $J$ [Eq.~\eqref{eq:afichain}] within AFI in (b) as $J = 4\gamma^2/U$.}\label{fig:fig5}
\end{figure}

Our {\em main} results in Figs.~\ref{fig:fig2}--\ref{fig:fig5} demonstrate how quantum STT deposits spin angular momentum [Figs.~\ref{fig:fig4} and ~\ref{fig:fig5}] into the AFMI by driving its on-site electronic spin expectation value from $\mathbf{s}^\prime_i(t=0) \equiv 0$ in equilibrium toward spatially inhomogeneous profile [Figs.~\ref{fig:fig2}(a) and ~\ref{fig:fig3}], $s_i^{\prime z}(t) \neq 0$ [$s_i^{\prime x}(t)=0=s_i^{\prime y}(t)$] with zigzag pattern $s_{2j-1}^{\prime z}(t) < s_{2j}^{\prime z}(t)$ for $j = 1, \ldots, N_{\mathrm{AFMI}}/2$.  The total  spin angular momentum absorbed by AFMI increases with the on-site Coulomb repulsion [Fig.~\ref{fig:fig5}(a)], but it is reduced [Figs.~\ref{fig:fig4}(c)] when the interchain hopping  allows for hybridization of NM and AFMI and electron leakage  from AFMI [Fig.~\ref{fig:fig4}(a)] into NM [Fig.~\ref{fig:fig4}(b)].  Prior to delving into these results, we introduce rigorous notation and useful concepts.

{\em Models and methods}.---The second-quantized many-electron Hamiltonian describing the NM/AFMI system in Fig.~\ref{fig:fig1}(a)  consists of four terms
\begin{equation}\label{eq:ham}
\hat{H} = \hat{H}_{\rm NM} + \hat{H}_{\rm AFMI} +  \hat{H}_{\rm NM-AFMI}+ \hat{H}_{\rm V, \mathbf{B}}(t < 0). 
\end{equation}
The first term is 1D TB Hamiltonian of noninteracting electrons within NM chain 
$
\hat{H}_{\rm NM} = -\gamma \sum_{i = 1}^{N} (
\hat{c}^\dagger_{i\uparrow}\hat{c}_{i+1\uparrow}
+
\hat{c}^\dagger_{i\downarrow}\hat{c}_{i+1\downarrow}
+
{\rm h.c.})
$
where $\hat{c}^\dagger_{i\sigma}$ ($\hat{c}_{i\sigma}$) creates (annihilates) an electron with spin $\sigma =
\uparrow, \downarrow$\ at site $i$, and $\gamma$ is the intrachain hopping. These operators act on four possible states at each site $i$---vacuum $|0\rangle$,  spin-up $|\!\!\uparrow\rangle$, spin-down  $|\!\!\downarrow\rangle$, and doubly occupied state $|\!\!\uparrow\downarrow\rangle$, so that total Hilbert space of NM/AFMI system has dimension $4^{92} \times 4^{12}$. The interacting electrons within the AFMI chain are described by the Hubbard Hamiltonian~\cite{Fradkin2013,Essler2005}
\begin{eqnarray}\label{eq:hmott}
\hat{H}_{\rm AFMI} & = & -\gamma \sum_{i = 1}^{N_{\rm AFMI}-1}
\left(
\hat{d}^\dagger_{i\uparrow}\hat{d}_{i+1\uparrow}
+
\hat{d}^\dagger_{i\downarrow}\hat{d}_{i+1\downarrow}
+
{\rm h.c.} \right) \nonumber \\
& & 
+U\sum_{i = 1}^{N_{\rm AFMI}}
\hat{n}'_{i\uparrow}\hat{n}'_{i\downarrow}.
\end{eqnarray}
Here, $\hat{n}'_{i\sigma} = \hat{d}^{\dagger}_{i\sigma} \hat{d}_{i\sigma}$ are electron density (per site) operator for spin $\sigma$ at site $i$ of AFMI. The on-site Coulomb repulsion, such as $U=0$--$10\gamma$ in Fig.~\ref{fig:fig3}(b), is expressed in the units of hopping $\gamma$ (typically \mbox{$\gamma=1$ eV}) which we use as a unit of energy. The operators for the total number of electrons, \mbox{$\hat{N}_e^\mathrm{AFMI}=\sum_i \hat{n}^\prime_{i}$}, and total spin along the $\alpha$-axis, \mbox{$\hat{s}^{\prime \alpha}=\sum_i \hat{s}^{\prime \alpha}_{i}$}, are given by the sum of electron and spin density operators, $\hat{n}^\prime_{i} = \sum_{\sigma=\{\uparrow,\downarrow\}}\hat{d}^\dagger_{i\sigma}\hat{d}_{i\sigma}$ and $\hat{s}^{\alpha \prime}_{i}=\sum_{\sigma=\{\uparrow,\downarrow\}} \hat{d}^\dagger_{i\sigma} \frac{1}{2}\hat{\sigma}^\alpha_{\sigma\sigma'} \hat{d}_{i\sigma'}$, respectively. The interchain exchange interaction $J_\mathrm{v}$ between electronic spins within NM and AFMI is described by 
\begin{eqnarray}\label{eq:int}
\lefteqn{\hat{H}_{\rm NM-AFMI}  = 
- J_{\rm v}\sum_{i = 1}^{N_{\rm AFMI}}
\hat{\mathbf{s}}_{i+N_{\rm
		L}}\cdot\hat{\mathbf{s}}'_{i}} \nonumber \\
&& \mbox{} -\gamma_{\rm v}\sum_{i = 1}^{N_{\rm AFMI}} 
\Big(
\hat{c}^\dagger_{i+N_{\textrm{L}}\uparrow}
\hat{d}_{i\uparrow}
+
\hat{c}^\dagger_{i+N_{\textrm{L}}\downarrow}
\hat{d}_{i\downarrow} + {\rm h.c.}
\Big),
\end{eqnarray}
where $\hat{\mathbf{s}}_{i}$ and $\hat{\mathbf{s}}^\prime_i$ are spin density (per site) operators in  NM and AFMI chains, respectively. Here we also add a term with possible $\gamma_{\rm v} \neq 0$ hopping between  $N_\mathrm{AFMI}$ sites of the central region of the NM chain and  $N_\mathrm{AFMI}$ sites of AFMI in Fig.~\ref{fig:fig1}(a), which can arise in realistic devices used in spintronics~\cite{Chen2018,Moriyama2018,Gray2019,Wang2019} due to evanescent wavefunctions.  They penetrate from the NM surface into the region of AFMI near the interface, thereby leading to charge transfer in equilibrium or current leakage between the two materials~\cite{Dolui2020}. Such normal-metal proximity effect on finite-size Mott insulators can also create exotic many-body states in equilibrium~\cite{Zenia2009}. To prepare the initial state of the conduction electrons in the NM chain, we confine them within  $N_\mathrm{conf}$ sites of the L lead in Fig.~\ref{fig:fig1}(a) and polarize their spins along the +$z$-axis by means of an additional term  $\hat{H}_{\rm V, \mathbf{B}}(t < 0)  = 
-V\sum_{i = 1}^{N_{\rm conf}}
\left(
\hat{c}^\dagger_{i\uparrow} 
\hat{c}_{i\uparrow}
+
\hat{c}^\dagger_{i\downarrow} 
\hat{c}_{i\downarrow}
\right)  -\sum_{i = 1}^{N_{\rm conf}} g\mu_{\rm B}
\hat{s}_{i}^z B_{\rm e}^z$.
Here  $V = 2\gamma$ is the confining potential; $B_{\rm e}^z$ is the external magnetic field; and $g\mu_{\rm B} B_{\rm e}^z= 10\gamma$, where $g$ is the electron gyromagnetic ratio and $\mu_{\rm B}$ is the Bohr magneton. After the initial state is prepared for $t<0$, 
$\hat{H}_{\rm V, \mathbf{B}}(t \ge 0)$ is set to zero, so that spin-polarized  electrons from the L lead 
propagate toward the R lead, as illustrated in Fig.~\ref{fig:fig1}(b) and computed in Fig.~\ref{fig:fig2}. 


In the limit $U \gg \gamma$, the half-filled ($n_i=1$) 1D Hubbard model describes electrons localized one per site, so it can be mapped~\cite{Fradkin2013,Essler2005} to isotropic AF quantum spin-$\frac{1}{2}$ Heisenberg chain with the effective Hamiltonian given in Eq.~\eqref{eq:afichain}. Therefore, for comparison we also analyze the NM/AFI setup  in Fig.~\ref{fig:fig1}(c) where AFI sites hosts localized spin-$\frac{1}{2}$ operators $\hat{\mathbf{S}}_i$, as described by the Hamiltonian $\hat{H} = \hat{H}_{\rm NM} + \hat{H}_{\rm AFI} +  \hat{H}_{\rm NM-AFI}+ \hat{H}_{\rm V, \mathbf{B}}(t < 0)$. Here $\hat{H}_{\rm NM}$ is the same as in  Eq.~\eqref{eq:ham}; $\hat{H}_{\rm AFI}$ is the same as in Eq.~\eqref{eq:afichain} where we use $J = 4\gamma^2/U$ as the exchange interaction in the limit $U \gg \gamma$~\cite{Fradkin2013,Essler2005}; the interchain interaction is described by  $\hat{H}_{\rm NM-AFI} = -J_{\rm v}\sum_{i = 1}^{N_{\rm AFI}} \hat{\mathbf{s}}_{i+N_{\textrm{L}}} \cdot  \hat{\mathbf{S}}_{i}$ where $J_{\rm v}=0.5\gamma$; and $\hat{H}_{\rm V, \mathbf{B}}(t < 0)$  is the same as in  Eq.~\eqref{eq:ham}. 

The tDMRG simulations~\cite{White2004,Feiguin2011,Daley2004,Paeckel2019} evolve the nonequilibrium state of the whole system in Fig.~\ref{fig:fig1}, \mbox{$|\Psi(t+\delta t) \rangle = e^{-i\hat{H} \delta t/\hbar} |\Psi(t)\rangle$}, using the time step \mbox{$\delta t =0.1 \hbar/\gamma$}. Additional 
details of tDMRG simulations are provided in the SM~\cite{sm}.


{\em Results and discussion}.---The Hubbard 1D chain modeling the AFMI possesses a sizable energy gap $\Delta_c$ for charge excitations at $U \gtrsim 2\gamma$, whose value is exactly known~\cite{Essler2005} in the limit $N_\mathrm{AFMI} \rightarrow \infty$ ($\Delta_c=0.173 \gamma$ at $U=2\gamma$; or $\Delta_c=0.631 \gamma$ at $U=3\gamma$). In chains of finite length, such as ours with $N_\mathrm{AFMI}=12$ sites, DMRG predicts slightly larger $\Delta_c$ values~\cite{Joost2020}. However, the spin sector of the half-filled Hubbard chain is gapless in the thermodynamic limit. This means that injecting a charge in the AFI is energetically costly, but creating a spin excitation is not. Figures~\ref{fig:fig2}(a) and ~\ref{fig:fig3} demonstrate that AFMI with  $U \gtrsim 4\gamma$ will be driven out of its GS with $\mathbf{s}_i^\prime =0$ on all sites toward a nonequilibrium phase with $s_i^{\prime z}(t) \neq 0$ and $s_i^{\prime x}=0=s_i^{\prime y}$ due to quantum STT exerted by injected current pulse  in the NM chain that is spin-polarized along the $z$-axis.  The spatial profile of $s_i^{\prime z}(t)$ is inhomogeneous with a zigzag pattern deep in the Mott insulator phase, which distinguishes it from the weak response of the borderline case with $U=2\gamma$ [Fig.~\ref{fig:fig3}(b)] or noninteracting chain with $U=0$  [Figs.~\ref{fig:fig2}(c) and ~\ref{fig:fig3}(b)]. 

Even after the current pulse in the NM chain has ended, the spin angular momentum remains deposited within AFMI, with 
its total value increasing with $U$ [Fig.~\ref{fig:fig5}(a)].  Such Mott insulator transmuted into a phase with nonzero total magnetization remains magnetized also when intrachain hopping is switched on, $\gamma_{\rm v}=0.1\gamma$, in Fig.~\ref{fig:fig4}(c). However, $\gamma_{\rm v}=0.1\gamma$ allows electrons to leak from AFMI [Fig.~\ref{fig:fig4}(a)] into NM [Fig.~\ref{fig:fig4}(b)] chain, so that total spin deposited into AFMI is reduced in Fig.~\ref{fig:fig4}(c) when compared to isolated AFMI. 

Figure~\ref{fig:fig5} explains quantum STT~\cite{Zholud2017,Mondal2019,Petrovic2020,Mitrofanov2020} as the transfer of total spin angular momentum from NM conduction electrons (dashed lines in Fig.~\ref{fig:fig5}) to confined electrons within the AFMI [solid lines in Fig.~\ref{fig:fig5}(a)] or to localized spins within the AFI [solid lines in Fig.~\ref{fig:fig5}(b)]. While some spin transfer exists even for $U=0$, it is dramatically enhanced by increasing $U$ to establish AFMI [Fig.~\ref{fig:fig5}(a)]. The NM/AFMI case with \mbox{$U=10\gamma$} shows that $\sum_i s_i^{\prime z}(t)$ within AFMI is nearly identical to $\sum_i S_i^z(t)$ within AFI with $J=4\gamma^2/U$, as anticipated from mapping~\cite{Fradkin2013,Essler2005} of AFMI to AFI in the limit $U \gg \gamma$. However, this correspondence fails for $U < 10 \gamma$. The absorbed spin by AFMI or AFI can be viewed as multiple excitations of any two-spinon or higher-order spinon states~\cite{Mourigal2013}, as long as they are compatible with total angular momentum conservation~\cite{Petrovic2020}.


{\em Conclusions.---} In conclusion, we demonstrate how tDMRG adapted~\cite{Petrovic2020} for {\em quantum STT}~\cite{Zholud2017,Mondal2019,Petrovic2020,Mitrofanov2020} makes it possible to study spin transfer into strongly electron-correlated antiferromagnets. 
In contrast,  quantum-classical theory of conventional STT~\cite{Ralph2008,Nunez2006,Haney2008,Xu2008,Stamenova2017,Gomonay2010,Hals2011,Cheng2014,Cheng2015,Saidaoui2014,Dolui2020} would conclude that 
entangled AF true GS does not undergo any current-driven dynamics when its localized spins have zero expectation value at $t=0$ as the initial state used in this study. Although tDMRG has been previously applied to study charge current through AFMI~\cite{Irino2010,Heidrich-Meisner2010,Dias2010} or spin-charge separation~\cite{Ulbricht2009} in geometries where electrons are injected into AFMI by finite bias voltage, spin-dependent transport phenomena in geometries like Fig.~\ref{fig:fig1} of relevance to spintronics~\cite{Chen2018,Moriyama2018,Gray2019,Wang2019} remain unexplored. Realistic spintronic devices would require to consider two- or three-dimensional geometries. But Keldysh Green functions~\cite{Gaury2014,Schlunzen2020}, as the only available nonequilibrium quantum many-body formalism  for higher dimensions and longer times, cannot at present access large $U$ with perturbative self-energies~\cite{Joost2020,Schlunzen2020}, or its nonperturbative implementation can handle~\cite{Bertrand2019} only a very few sites. Therefore, this study represents a pivotal test case that provides intuition about quantum STT phenomena in strongly correlated and/or entangled quantum materials, as well as a benchmark~\cite{Schlunzen2020} for any future developments via the Keldysh Green functions.

\begin{acknowledgments}
M.~D.~P., P.~M. and  B.~K.~N. were supported by the U.S. National Science Foundation (NSF) Grant No. ECCS 1922689. A.~E.~F. was supported by the U.S.  Department of Energy (DOE) Grant No.~DE-SC0019275.
\end{acknowledgments}

\end{document}